\documentstyle[epsf]{article}
\begin{document}
\title{Implementation of the Deutsch-Jozsa algorithm with Josephson 
       charge qubits}
\author{Jens Siewert$^{1,2}$ and Rosario Fazio$^{2,3}$\\*[2mm]
\footnotesize\it
        $^1$      Institut f\"ur Theoretische Physik, 
                   Universit\"at Regensburg,
                   93040 Regensburg, Germany\\
\footnotesize\it
        $^2$      Dipartimento di Metodologie Fisiche e Chimiche (DMFCI), 
	           Universit\`a di Catania,\\
\footnotesize\it
		   viale A.Doria 6, I-95125 Catania, Italy, and\\
\footnotesize\it
	           Istituto Nazionale per la Fisica della Materia (INFM),
	           Unit\`a di Catania\\
\footnotesize\it
        $^3$      National Enterprise for Nanoscience and Nanotechnology
	           (NEST) - INFM,\\
\footnotesize\it
		   and Scuola Normale Superiore, I-56126 Pisa, Italy}
\normalsize
\rm
\date{\today}
\maketitle
\begin{abstract}
        We investigate the realization of a simple solid-state
        quantum computer by implementing 
        the Deutsch-Jozsa algorithm in a system of Josephson
        charge qubits.
        Starting from a procedure to carry out the 
        one-qubit Deutsch-Jozsa algorithm we show
        how the $N$-qubit version of the
        Bernstein-Vazirani algorithm can be realized.
        For the implementation of the three-qubit Deutsch-Jozsa algorithm 
        we study in detail a setup which allows to produce entangled states.
\end{abstract}
\vspace*{2mm}
\\[5mm]
\section{Introduction}
The growing interest in quantum computation has stimulated 
an impressive activity in the field of `quantum hardware'. 
Numerous proposals to implement quantum bits and simple one-bit and
two-bit operations
have appeared in many different areas of contemporary physics research.
Yet the possibility to taylor a controllable two-state system is by
far not enough to do quantum computation. 
From an engineering point of view the design
of a quantum computer is difficult because of the enormous complexity
of requirements, such as the possibility to prepare and to measure states
easily, a highly flexible setup with a sufficiently large parameter space
that can be controlled, the maintainance of coherence etc.  
Therefore, the touchstone of practical quantum computation is the
implementation of quantum algorithms. 

This goal might appear too ambitious, in particular if one thinks
of realizing a set of universal gates \cite{UniversalQC},  i.e.\
a quantum computer which can do \underline{any} operation. 
However, in order to implement \underline{one particular} 
quantum algorithm it is
not necessary to go as far as this. Since usually one has to deal
with a well-defined set of input and output states it suffices to
implement just those gates which represent the desired operations on 
applying the gate to the possible input states. 
This renders the task easier than designing  gates
which represent the operations on application to \underline{any} state.

Another advantage of the restriction to a particular algorithm 
is the possibility to use more complex operations
(for example $N$-bit gates instead of a sequence of one-bit and
two-bit operations) which require less computational time. This could be
one way to overcome the limits which are set by a small decoherence
time in a given physical system.

Surprisingly, there has been comparatively little activity 
towards the implementation of quantum algorithms in real physical
systems.  To date, quantum algorithms  have been implemented 
by using liquid-state NMR 
\cite{Cory97,Gershenfeld97,Chuangetal98,JonesMoscaHansen,Linden98,%
      Arvind00,Arvind99,Collins00}, 
in atomic physics \cite{Ahn2000} and by
optical interferometry \cite{Kwiat99}.
A solid-state implementation
of Grover's algorithm has been proposed theoretically \cite{Loss2000}.

The quest for large scale integrability has stimulated 
an increasing interest in  superconducting 
nanocircuits \cite{ShnirmanPRL,Averin98,Makhlin99,Caspar99,nature}
as possible candidates for the implementation of a quantum computer.
The recent experimental breakthrough for Josephson 
qubits~\cite{Nakamura99,Friedman,Caspar00} is the first important step 
towards a solid-state quantum computer.

Naturally the question arises whether, at the present
technological level, it is possible 
to implement also quantum algorithms in these systems. 
Here we concentrate on charge 
qubits~\cite{ShnirmanPRL,Averin98,Makhlin99} and show how the 
Deutsch-Jozsa algorithm \cite{Deutsch85,Deutsch92,Cleve98,Collins98} and the 
Bernstein-Vazirani algorithm \cite{Bernstein93} can be run on a 
Josephson quantum computer. 
We analyse the experiment by Nakamura et al.\ \cite{Nakamura99}
in terms of quantum interferometry \cite{Cleve98} and show 
that it corresponds to the implementation of the one-qubit version
of Deutsch's algorithm. By generalising this idea we show how 
the $N$-qubit Deutsch algorithm, with $N\leq3$, can be implemented.

\section{Deutsch-Jozsa algorithm}
\label{section2}
Consider the subset of Boolean functions $f:\ \{0,1\}^N \rightarrow \{0,1\}$ 
with the property that $f$ is either constant or balanced (that is,
it has an equal number of 0 outputs as 1s). 
The Deutsch-Jozsa 
algorithm \cite{Deutsch85,Deutsch92,Cleve98} determines 
-- for a given unknown function $f$ -- 
whether the function is 
constant or balanced. 
With a classical algorithm, this problem would, in
the worst case, require $2^{N-1}+1$ evaluations of $f$ whereas the
quantum algorithm solves it with a single evaluation by means of the following 
steps (here we focus on the  refined version 
by Collins et al.\ \cite{Collins98},
see also Fig.\ 1). 
\newline\indent
{\em (i)} All qubits are prepared in the initial state $|0\rangle$,
therefore the $N$-qubit register is in the state $|00\ldots 0\rangle$.
\newline\indent
{\em (ii)} Perform an $N$-qubit Hadamard  
transformation $\cal H$ 
\begin{equation}
|x\rangle \ \stackrel{\cal H}{\longrightarrow} \sum_{y\in \{0,1\}^N}
                                             (-1)^{x\cdot y}\ |y\rangle\ , 
           \ \ \ (x\in \{0,1\}^N),
\label{HadamardTrafo}
\nonumber
\end{equation}
where
$
     x\cdot y = (x_1 \wedge y_1) \oplus \ldots \oplus (x_N \wedge y_N)
$
is the scalar product modulo two. This is equivalent to 
performing a one-bit Hadamard transformation to each qubit individually.
\newline\indent
{\em (iii)} Apply the $f$-controlled phase shift $U_f$ 
\cite{Cleve98,Collins98}
\begin{equation}
    |x\rangle\ \stackrel{U_f}{\longrightarrow}\ (-1)^{f(x)}\ |x\rangle\ ,\ \ \
        (x\in \{0,1\}^N)\ .
\label{f-phaseshift}
\end{equation}
\newline\indent
{\em (iv)} Perform another Hadamard transformation $\cal H$. 
\newline\indent
{\em (v)} Measure the final state of the register. If the result
is $|00\ldots 0\rangle$ the function $f$ is constant; if, however,
 the amplitude $a_{|00\ldots0\rangle}$ of the state 
$|00\ldots 0\rangle$ is zero the function $f$ is balanced.
This is because 
\begin{equation}
     a_{|00\ldots 0\rangle} = \frac{1}{2^N}
                                 \sum_{x\in \{0,1\}^N} (-1)^{f(x)}\ \ .
\label{amplitude0}
\end{equation}

We note that it is reasonable to identify functions $f$ whose outputs
can be mapped into each other by a bitwise NOT. For these functions, 
the gate operations in Eq.\ (\ref{f-phaseshift}) differ
merely by a global phase factor $(-1)$. We will use the convention
$f(00\ldots0)=0$. This reduces the number of gates which need to be
implemented, by a factor of two and does not affect the exponential 
speed-up.

In order to implement the algorithm it is necessary to show that each
individual step (preparation of the state, gate operations, measurement)
can be realized in a given system. It is well known how to prepare
and to measure the states in Josephson charge qubits 
\cite{ShnirmanPRL,Makhlin99,Nakamura99}. Our task is to demonstrate
that the gate operations corresponding to \underline{all} admissible functions
$f$ can be performed with a single device. An important aspect of our
proposal is that the gate operations are represented in a basis of
superpositions of charge states.

It is interesting to note that according to Ref.\ \cite{Collins98} 
the Deutsch-Jozsa algorithm does not involve entanglement for $N\le 2$, 
i.e.\  the $N=2$ version can be realized with uncoupled qubits.
On the other hand, the implementation of the algorithm for $N>2$ 
involves entanglement~\cite{Azuma01} and therefore requires a setup 
of coupled qubits.

\section{Bernstein-Vazirani algorithm}
The Bernstein-Vazirani algorithm \cite{Bernstein93,Cleve98}
is analogous to the $N$-bit Deutsch-Jozsa algorithm described
in the previous section, with the difference that 
the function $f$ has the form 
\begin{equation}
 f = a\cdot x\oplus b\ \ ,\ \ (a,x\in \{0,1\}^N,\ b\in \{0,1\})\ \ .
\label{f-BV}
\end{equation}
The gate in step {\em (iii)} is denoted by $U_a$ and assumes the form
\begin{equation}
|x\rangle \ \stackrel{U_a}{\longrightarrow} 
                            (-1)^{a\cdot x\oplus b}\ |x\rangle \ \ .
\label{Ua-BV}
\end{equation}
By measuring the register after running the algorithm once
one obtains the number $a$ in binary representation which
classically would require $N$ function calls.
The Bernstein-Vazirani
algorithm does \underline{not} involve entangled states at all \cite{Meyer}. 
This can be seen explicitely by rewriting the gate $U_a$ as a product 
of single-qubit gates 
\begin{equation} 
 U_a = (-1)^b \ \prod_{j=1}^N (\sigma_z^{(j)})^{a_j}
\label{UaOp}
\end{equation} 
where $a_j$ denotes the $j$th digit of $a$ in binary representation.
Here we have used the Pauli operators $\sigma^{(j)}_k$ acting on the
computational basis of the $j$th qubit $\{|0_j\rangle, |1_j\rangle\}$ and
the definition $(\sigma_k)^0 :=: I_1$ (with the one-qubit 
identity operator $I_1$). In particular we choose 
$\sigma_z^{(j)}\ |0_j\rangle = +|0_j\rangle$, 
$\sigma_z^{(j)}\ |1_j\rangle = -|1_j\rangle$.

Apart from the global phase $(-1)^b$ the set of gates $U_a$
represents the part of the Deutsch
algorithm with completely separable gates. As the algorithm starts with a 
product state, no entanglement is involved at any step. 
We note that by rewriting the action of the
whole algorithm as
\[
{\cal H}U_a{\cal H}\ |00\ldots0\rangle\ =\ 
(-1)^b \prod_j (\sigma_x^{(j)})^{a_j}\ |0_j\rangle\ =\
  (-1)^b  \prod_j |a_j\rangle
\] 
one sees that it leads trivially to the result.
In conclusion, it is possible to realize the Bernstein-Vazirani algorithm 
using uncoupled qubits in complete analogy with the implementation
for the one-qubit and two-qubit Deutsch algorithm.

\section{Josephson charge qubits and implementation of algorithms 
         without entanglement}
A Josephson charge qubit \cite{ShnirmanPRL,Makhlin99}
is a Cooper-pair box (see Fig.\ 3a) 
which is characterized by two energy scales, the charging energy 
$E_{\rm ch}=(2e)^2/(2C)$ (here $C$ is the total capacitance of
the island) and the Josephson energy $E_J\ll E_{\rm ch}$ 
of the tunnel junction. 
At low temperatures $T\ll E_J$ 
only the two charges states with 0 and 1 excess
Cooper pair on the island are important and 
the system behaves as a two-level system with the Hilbert space 
$\{|0\rangle,|1\rangle \}$ and the one-qubit Hamiltonian
\begin{equation}
    H_{\rm 1q}= (E_{\rm ch}/2)\ (2n_x-1)\ \sigma_z 
                           \    -\ (E_J/2)\ \sigma_x \ \ .
\label{1qHamiltonian}
\end{equation}
Here $n_x=C_g V_g/(2e)$ is the offset charge which can 
be controlled by the gate voltage. 

In a recent experiment, Nakamura et al.\ have measured
Rabi oscillations in a Josephson charge qubit \cite{Nakamura99}. 
In the following we analyse 
briefly the experiment and argue that it can be regarded
as the one-qubit implementation of Deutsch's algorithm.

The experiment is shown schematically in Fig.\ 2.
Let us first consider the part between the two sudden switchings of
the gate voltage. The system has been prepared in a superposition of
the two states $| + \rangle$, $| - \rangle$ which are defined as
\begin{equation}
        |\pm \rangle\ = \ \frac{1}{\sqrt{2}}\ 
	                   (| 0 \rangle\ \pm\ | 1 \rangle)\ \ .
\label{diagbasis}
\end{equation}
Now a rotation 
\begin{equation}
\exp{(i(E_Jt/2\hbar)\sigma_z)}  
\label{rabirot}
\end{equation}
is performed on the state which results in the final state 
\[ \frac{1}{\sqrt{2}}(\ |+\rangle\  +\
               e^{-i E_J t/\hbar}\ |-\rangle\ )\ \ .
\]
Note that here $\sigma_z$ is the Pauli operator with respect to the 
basis $\{|+\rangle,|-\rangle\}$. 

After sweeping back the gate voltage, a \underline{charge} 
measurement is performed. 
That is, the final superposition is measured in the charge basis 
$\{|0\rangle, |1\rangle\}$. In particular, for $t=2\pi\hbar/E_J$
the outcome of an ideal measurement is $|0\rangle$ while for
$t=\pi\hbar/E_J$ the state $|1\rangle$ is found. 

In a spin-related language the experiment could be described as follows.
The charge states form the $z$ basis, the charging energy corresponds
to the magnetic field in $z$ direction, and the Josephson energy
corresponds to the field in $x$ direction 
(cf.\ Eq.\ (\ref{1qHamiltonian})). The state is prepared 
with the magnetic field in $z$ direction. Switching the gate voltage 
suddenly to
the degeneracy point flips the magnetic field to the $x$
direction. Thus, the spin starts to precess around the $x$ axis.
After the operation time $t$ the field in $z$ direction is switched
on again, thus freezing the $z$ component of the spin. 
The latter corresponds to the island charge which then is
measured.

In order to display the analogy between Deutsch's algorithm 
(see Section \ref{section2}) 
for one qubit and Nakamura's experiment, we rephrase the algorithm
in the following way.
In step {\it (i)} and {\it (ii)} the symmetric superposition of
the states of the computational basis $\{|+\rangle, |-\rangle\}$ 
is prepared. On applying $U_f$ in step {\it (iii)}
the sign of $|-\rangle$ in the superposition is changed
if the function $f$ is balanced, or it is left unchanged. 
The purpose of the second Hadamard gate is 
the transformation of the superposition to a pure state which is
to be measured.

Now compare this sequence to Nakamura's experiment.
It starts with the preparation of the symmetric
superposition. The Rabi oscillation corresponds to the application of
the gate $U_f$ which is implemented by choosing the operation time 
appropriately (see Table I).
%
%
Instead of performing the second Hadamard gate the superposition
is \underline{measured directly}. This is possible because each superposition
corresponds to a charge eigenstate: if $f$ is constant, the charge state
$|0\rangle$ is obtained while balanced $f$ leads to charge state $|1\rangle$.

The additional advantage of this procedure is that the second sudden
sweep brings the Hamiltonian back to a regime where the charge states
are approximately the eigenstates such that the island charge is frozen.
This is important since the time of the measurement is much bigger than
the intrinsic time scale $\hbar/E_J$ of the qubit.

Thus, Nakamura's experiment with the operation times chosen 
according to Table I
can be viewed as the implementation of the one-bit Deutsch algorithm.
The only difference compared to the sequence of steps in Section II is
that the single-qubit superpositions are prepared and measured directly, without
performing additional one-qubit operations \cite{footnote}.
Of course, the essence of the algorithm is not affected: instead of rotating
the state `forward' by a Hadamard operation we apply a Hadamard 
rotation `backwards' to the basis. It is easily seen that the 
coefficient $a_{|00\ldots0\rangle}$ of the \underline{charge} state 
$|00\ldots0\rangle$ still obeys Eq.\ (\ref{amplitude0}) where now
the sum has to be taken over all $x\in \{+,-\}^N$.

The two-qubit Deutsch-Jozsa algorithm can be 
implemented by using two uncoupled qubits \cite{Collins98}. 
The gates 
$\sigma_z^{(1)} \otimes I_1^{(2)}$, $I_1^{(1)} \otimes \sigma_z^{(2)}$,
$\sigma_z^{(1)}\otimes \sigma_z^{(2)}$ implementing the balanced functions
(the upper index denotes the qubit number) and
$I_1^{(1)}\otimes I_1^{(2)}$ for the constant function can be realized
in complete analogy to the one-qubit algorithm.
If, on measuring the qubits, both of them
are found in the charge state $|0\rangle$ the function $f$ was constant;
if at least one qubit is found in the state $|1\rangle$, $f$ was balanced.

Finally, it is also obvious that the Bernstein-Vazirani algorithm
can be realized by applying the procedure described above.
According to Eq.\ (\ref{UaOp}) one needs to implement the operation
$(\sigma^{(j)}_z)^{a_j}$ for the $j$th qubit which is straightforward
by using the entries of Table I (note that the number $b$ in Eq.\ 
(\ref{UaOp}) is irrelevant for the implementation). 
The measurement of the register then
yields precisely the binary representation of $a$.

\section{Implementation of algorithms involving entanglement -- the 
         three-qubit Deutsch-Jozsa algorithm}
The realization of the three-qubit version of the algorithm is 
more difficult. 
Apart from the constant function 
35 balanced functions need to be implemented. Moreover,
the three-qubit algorithm involves gates $U_f$ which produce
entangled final states. 

The goal is to proceed along the same lines as above, that is, 
preparation of the state
$|000\rangle$, sudden sweep of $n_x^{(j)}$ etc.
The action of the gates $U_f$ takes place
in the basis \mbox{$\{|+++\rangle, |++-\rangle,\ldots,|---\rangle\}$.}
That is, the gates operate at the degeneracy point $n_x^{(j)}=1/2$
of the charge qubits. 
In order to find 
efficient ways for the implementation we first analyse the
functions $f$ and the corresponding gates $U_f$.

Apart from the constant function and its gate 
$I^{(1)}_1\otimes I^{(2)}_1\otimes I^{(3)}_1$ there are 7 balanced
functions for which the gates are separable: 
$\sigma^{(1)}_z \otimes I^{(2)}_1\otimes I^{(3)}_1$,
$I^{(1)}_1\otimes \sigma^{(2)}_z\otimes I^{(3)}_1$, 
\ldots,
$\sigma_z^{(1)}\otimes \sigma_z^{(2)}\otimes \sigma_z^{(3)}$.
Further there are 12 balanced functions for which the 
gates factorize into a one-qubit part and a two-qubit
part as in example $I)$ below. 
The other gates entangle all three qubits and can be divided
into two classes (see example $II)$ and $III)$).
There are 12 gates of class $II)$ and 4 gates of class $III)$.
\[
\begin{array}{ll}
I) &
            \displaystyle\frac{1}{2}\
     \left(    I^{(1)}_1\otimes I^{(2)}_1
             + \sigma^{(1)}_z\otimes I^{(2)}_1   
            - I^{(1)}_1\otimes \sigma^{(2)}_z
            + \sigma^{(1)}_z\otimes \sigma^{(2)}_z
     \right)
              \otimes \sigma_z^{(3)}
\\[4mm]
II)  &
            \displaystyle\frac{1}{2}
            \left( \sigma_z^{(1)} \otimes I^{(2)}_1\otimes I^{(3)}_1 
        -  I^{(1)}_1\otimes I^{(2)}_1 \otimes \sigma_z^{(3)} 
             +  \sigma_z^{(1)} \otimes \sigma^{(2)}_z\otimes I^{(3)}_1\ +
	     \right.
        \\ &\hfill \left.
	 + \  I^{(1)}_1\otimes \sigma^{(2)}_z \otimes \sigma_z^{(3)}
            \right)  
 \\[4mm]
III) &
            \displaystyle\frac{1}{2}
            \left( \sigma_z^{(1)} \otimes I^{(2)}_1\otimes I^{(3)}_1 
        -  I^{(1)}_1\otimes \sigma^{(2)}_z \otimes I_1^{(3)} 
                +  I_1^{(1)} \otimes I^{(2)}_1\otimes \sigma^{(3)}_z\ +
	     \right.
        \\ &\hfill \left.
            +\  \sigma^{(1)}_z\otimes \sigma^{(2)}_z \otimes \sigma_z^{(3)}
            \right)
\end{array}
\]
All separable single-qubit operations can be carried out in analogy 
with the one-qubit case above. In the following we discuss
how the entangling gate operations can be achieved.
For the realization of these gates a coupling of tunable strength
between the qubits is required. 

There are 
various ways to couple charge qubits 
\cite{ShnirmanPRL,Averin98,Makhlin99,nature}. 
Here we investigate 
coupling via
Josephson junctions \cite{ourJLTP}. 
Each qubit island is coupled to its nearest neighbour
using a symmetric SQUID
(see Fig.\ 3b). 
%
%
%
%

Assuming that both the $j$th qubit and the $j$th
coupling junction are tunable by local fluxes $\Phi^{(j)}$, $\Phi^{(j)}_K$
the Hamiltonian for the $N$-qubit system at 
the degeneracy point $n_x^{(j)}=1/2$ reads
\begin{equation}
  \begin{array}{cl}
    H_{N{\rm q}} & = \displaystyle
                     \sum_{j=1}^N \ \ 
                                  \left\{ 
                                 H_{\rm 1q}^{(j)}(\Phi^{(j)}) 
   \    +\   E_K^{(j)}\  \sigma_z^{(j)}\sigma_z^{(j+1)}
                                  \right.
              \\[4mm] &           \left.
              -\ (1/2)\ J_K^{(j)}(\Phi_K^{(j)})\ 
                 [\ \sigma^{(j)}_+ \sigma^{(j+1)}_- \ +\ \ {\rm h.c.\/}\ ]
                                \right\}\ \ .
  \end{array}
\label{NqHamiltonian}
\end{equation}
Here $J_K^{(j)}$ is the Josephson energy of the $j$th coupling SQUID
and $\sigma_\pm=(\sigma_x\pm i\sigma_y)/2$. In the limit
of small coupling capacitances
$C_K^{(j)}\ll C^{(j)}$ we have
\[
       E_K^{(j)}=\frac{C_K^{(j)}}{2C^{(j)}} E_{\rm ch}^{(j)}\ \ .
\]
We will assume that $E_K^{(j)}$ is negligible.
Since in practise the capacitive coupling is
always present it is necessary to have $J_K^{(j)}(\Phi^{(j)}_K=0)
                                                            \gg 4E_K^{(j)}$.
Then the dynamics of the system approximates the ideal dynamics 
sufficiently well. 

Consider now the first and the second qubit coupled by $J_K^{(1)}$.
By choosing,  e.g.\ 
$-E_J^{(1)}=E_J^{(2)}=\pm J_K^{(1)}={\sf J}$ and the operation
time $t\simeq0.97 (2\pi/{\sf J})$ we obtain 
an operation similar to a swap gate for the qubits 1 and 2
for which we introduce the notation
(in the basis $\{|++\rangle,\ldots,
                   |--\rangle \}_{12}$) 
\begin{equation}
        [\pm 12] :=: \left(
                              \begin{array}{cccc}
                                0 & 0 & 0 & \pm i \\
                                0 & 1 & 0 & 0 \\
                                0 & 0 & 1 & 0 \\
                                \pm i & 0 & 0 & 0 
                              \end{array}
                   \right) \ \ .
\end{equation}
By denoting one-bit phase shifts for the $j$th qubit 
\begin{equation}
        [\pm j] :=: \left(
                              \begin{array}{cc}
                                1 & 0 \\
                                0 & \pm i 
                              \end{array}
                   \right) \ \ ,
\end{equation}
we can write a sequence of operations which gives the two-bit
entangling gate in example $I)$ above:
\begin{equation}
                       [+1][-2] \frac{\rule{.4cm}{.0cm}}{\rule{.4cm}{.0cm}}
                       [-12] 
   \frac{\rule{.4cm}{.0cm}}{\rule{.4cm}{.0cm}}   \sigma^{(3)}_z   
                       \ \ .
\label{exI}
\end{equation}
After suddenly sweeping $n_x^{(1)}$ and $n_x^{(2)}$ to the degeneracy,
first the one-bit phase shifts are performed by keeping $J_K^{(1)}=0$.
Then $J_K^{(1)}$ is switched on suddenly in order to do the
two-bit rotation. The $\sigma^{(3)}_z$ rotation can be done at any moment
since the third qubit is decoupled from the other two. Finally
the $n_x^{(j)}$ are swept back suddenly and the register is measured.
Note that the parameters of the one-bit and two-bit operations 
need to be chosen in a compatible
way, i.e.\ the local Josephson couplings $E_J^{(1,2)}$ should be the same
for the one-bit and two-bit operations in order to avoid unnecessary parameter
switching. 

There are numerous ways to represent the three-bit entangling gates. 
At least two different two-bit rotations need to be applied.
During the second two-bit rotation the third qubit has to be `halted'.
This can be done by switching off both the $E_J$ and the $J_K$ which
couple to this qubit. 
A possible sequence for example $II)$ is
\begin{equation}
                       [+1][-2] \frac{\rule{.4cm}{.0cm}}{\rule{.4cm}{.0cm}}
                       [+13] \frac{\rule{.4cm}{.0cm}}{\rule{.4cm}{.0cm}}
                       [-12] 
                       \ \ ,
\label{exII}
\end{equation}
and for example $III)$
\begin{equation}
 \sigma_z^{(1)}\sigma_z^{(2)}\sigma_z^{(3)}
                       \frac{\rule{.4cm}{.0cm}}{\rule{.4cm}{.0cm}}
 [+12]                 \frac{\rule{.4cm}{.0cm}}{\rule{.4cm}{.0cm}}
 [+23]                 \frac{\rule{.4cm}{.0cm}}{\rule{.4cm}{.0cm}}
 [+12] \ \ .
\label{exIII}
\end{equation}

It is interesting to note that the completely entangling gates 
of class $II)$  and $III)$ can be realized approximately
with a \underline{single} three-qubit operation.
In Table II and III we list the parameters for the various implementations
 including estimates for the accuracy of the respective operation.
%
%
%
%
The complete set of entangling gates can be obtained from the sequences 
(\ref{exI}) -- (\ref{exIII})
by cyclic permutations of qubit numbers (and appropriate sign changes),
thereby paying attention that the parameter settings are compatible for
both one-bit and two-bit operations.

\section{Conclusions}
We have presented a possible implementation of the Deutsch-Jozsa algorithm
in a setup of Josephson charge qubits. While this algorithm for a qubit
number $N\ge 3$ requires entanglement we have demonstrated explicitely
that the Bernstein-Vazirani algorithm does not involve entanglement for
any number of qubits. 

Our implementation realizes the algorithms by using only state-of-the-art
technology. A peculiarity is that the gate operations representing the
algorithm  are carried out in a basis which is different from the one
which is measured. This helps us to obtain the desired results 
with a minimum number of operations. Thus, one may hope to see
the expected behaviour of the system even with the relatively small
decoherence times which have been achieved up to now \cite{Nakamura99}. 
Of course, these
measured decoherence times refer to a single qubit; at the moment
it is not clear how much more difficult it is to observe
entangled charge qubits experimentally. In fact, the experimental 
implementation of this proposal may serve to study this question in detail.

Finally, we mention that the methods outlined above can 
be used to study also other interesting problems 
such as the production and measurement of Bell states and GHZ states
\cite{saroprep}.
From a practical point of view, it would be particularly interesting to 
find ways to create such states in a `single shot' with one appropriate 
gate operation. Even though it appears
rather difficult to avoid the locality loophole in this kind of setup it
is nevertheless a remarkable challenge to measure such quantum 
correlations in a  macroscopic system.

\section{Acknowledgements}
The authors would like to thank L.\ Amico, D.V.\ Averin,  I.\ Chuang, 
P.\ Delsing, G.\ Falci, J.B.\ Majer,
Y.\ Makhlin, A.\ Osterloh, G.M.\ Palma, F.\ Plastina, 
C.\ Urbina, V.\ Vedral and C.\ v.d.\ Wal 
for stimulating discussions. 
This work was supported in part by the EC-TMR, IST-Squbit
and INFM-PRA-SSQI.

\newpage
%
%
%
%
%
%
%
\begin{table} 
\begin{tabular}{|c|c|c|}\hline
     f     & \parbox{1cm}{  \begin{center}
                              gate $U_f$
                            \end{center}  }  
           & \parbox{1cm}{  \begin{center}
                              time $t$ 
                            \end{center}  }     \\ \hline
     constant & $I_1$         & $2\pi\hbar/E_J$ \\ \hline
     balanced & $\sigma_z$    & $\pi\hbar/E_J$ \\ \hline 
\end{tabular} 
\caption{}
\end{table} 
\vspace*{10cm}
\newpage
\begin{table}
\begin{tabular}{|c|c|c|c|c|c|c|c|}\hline
  gate & implementation & $E_J^{(1)}$ & $E_J^{(2)}$ & $E_J^{(3)}$    
                        & $J_K^{(1)}$ & $J_K^{(2)}$ & $J_K^{(3)}$    
                        \\ \hline\hline
 $II)$ & sequence (9)   & {\sf -J} & {\sf J} & {\sf J} 
                        & {\sf -J} &  0      & {\sf J} 
                        \\ \hline
 $II)$ & single operation   & {\sf -J/2} & 0       & {\sf J}/2 
                        & {\sf J} & {\sf J} & 0
                        \\ \hline
$III)$ & sequence (10)  & {\sf J} & {\sf -J}& {\sf J} 
                        & {\sf J} & {\sf J} & 0 
                        \\ \hline
$III)$ & single operation   & {\sf J}/2 & {\sf -J}/2 & $0.83 {\sf J}$ 
                        & {\sf 0} & {\sf J} & {\sf J} 
                        \\ \hline
\end{tabular} 
\caption{Junction parameters for various realizations of the gates $II)$ and
         $III)$.}
\end{table}  
\vspace*{10cm}
\newpage
\begin{table}
\begin{tabular}{|c|c|c|c|c|}\hline
  gate & implementation 
                        & time $t/(2\pi\hbar/{\sf J})$ 
                        & \parbox{2.cm}{\begin{center}
                                             $a_{|0\ldots0\rangle}$\\
                                             $(E_K^{(j)}=0)$
                                        \end{center}} 
                        & \parbox{2.3cm}{\begin{center}
                                             $a_{|0\ldots0\rangle}$\\
                                             $(E_K^{(j)}={\sf J}/40)$
                                        \end{center}} 
                        \\ \hline\hline
 $II)$ & sequence (9)   
                        & 0.97 (2bit op.\/)
                        &  2 $\cdot$ $10^{-3}$     & 2 $\cdot$ $10^{-4}$ 
                        \\ \hline
 $II)$ & single operation 
                        & 0.80
                        & 7 $\cdot$ $10^{-5}$     &  2 $\cdot$ $10^{-2}$
                        \\ \hline
$III)$ & sequence (10)  
                        & 0.97 (2bit op.\/)
                        & 3 $\cdot$ $10^{-4}$     &  6 $\cdot$ $10^{-3}$
                        \\ \hline
$III)$ & single operation  
                        & 1.19
                        & $< 10^{-5}$     &  2 $\cdot$ $10^{-3}$
                        \\ \hline
\end{tabular} 
\caption{Gate operation times and
         accuracy for gate realizations of examples $II)$ and $III)$.
         The coefficient $a_{|0\ldots0\rangle}$ is a measure for the 
         fidelity of the operation (for an ideal operation
         $\ a_{|0\ldots0\rangle} = 0$, cf.\ Eq.\
	 (\protect\ref{amplitude0})). The operation time for the sequences
         refers to the time needed for the two-qubit rotations.
         Single-qubit rotations are assumed to be perfect.}
\end{table}  
\vspace*{10cm}

\newpage
\begin{figure}
\vspace*{3cm}
\caption{The sequence of operations to perform the Deutsch algorithm 
         on a register of $N$ qubits. According to Ref.\
         \protect\cite{Cleve98} it  can be interpreted in terms 
         of quantum interferometry. The first Hadamard transformation
         produces a superposition of \underline{all} possible states.
         Thus, with the application of the $f$-controlled gate $U_f$
         the outcome of $f$ for all possible arguments is evaluated
         at the same time.
         The second Hadamard transformation brings all computational
         paths together.} 
\end{figure}
\begin{figure}
\vspace{3cm}
\caption{The qubit is prepared in the ground state $|0\rangle$.
         After suddenly sweeping the gate voltage the system starts 
         Rabi oscillations between the eigenstates of the new
         Hamiltonian \protect$|\pm\rangle = (|0\rangle\pm |1\rangle)/\sqrt{2}$.
         After the time $t$ the gate voltage is swept back suddenly
           which freezes the final state; then the qubit is measured.} 
\end{figure}

\begin{figure}
\vspace{3cm}
\caption{a) A charge qubit. The Josephson energy of the junction
         can be controlled by the magnetic flux $\Phi$:
         $E_J(\Phi)=2{\cal E}_J \cos{(\pi\Phi/\Phi_0)}$, where
         ${\cal E}_J$ is the Josephson energy of the junctions
         of the symmetric SQUID and $\Phi_0=h/(2e)$ is
         the flux quantum \protect\cite{Makhlin99}.
         b) A possible realization of coupled charge qubits.}
\end{figure}

\begin{thebibliography}{99}
\bibitem{footnote} In Section \ref{section2}, the Hadamard transformations of
   the steps {\it (ii)} and {\it (iv)}  are present only for 
   `technical reasons': it is assumed that the basis states on which the gate 
   $U_f$ acts (the computational basis)
   are the same states which can easily be prepared and measured.
\end{thebibliography}
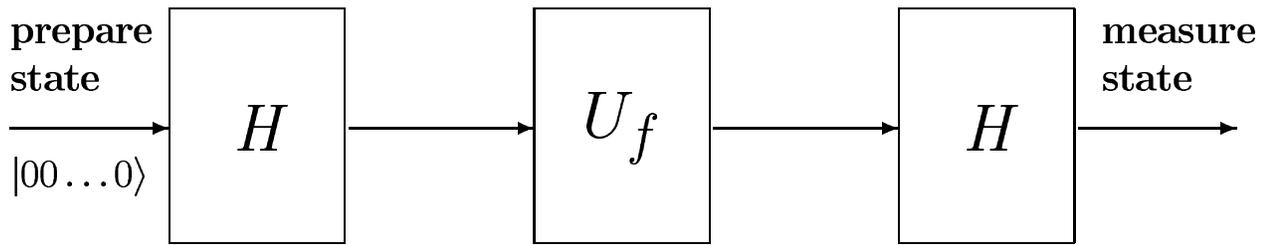
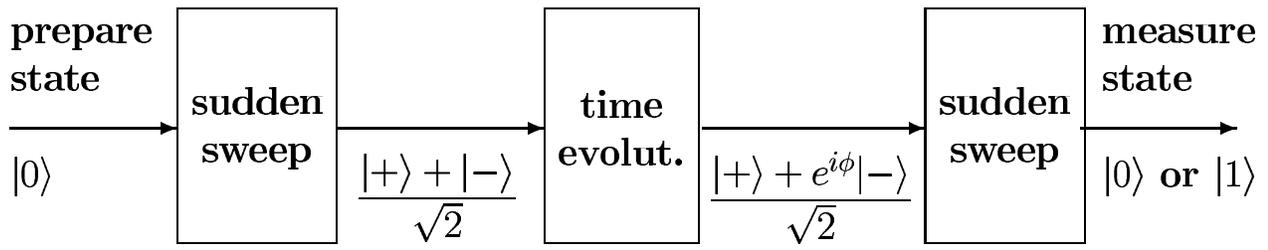
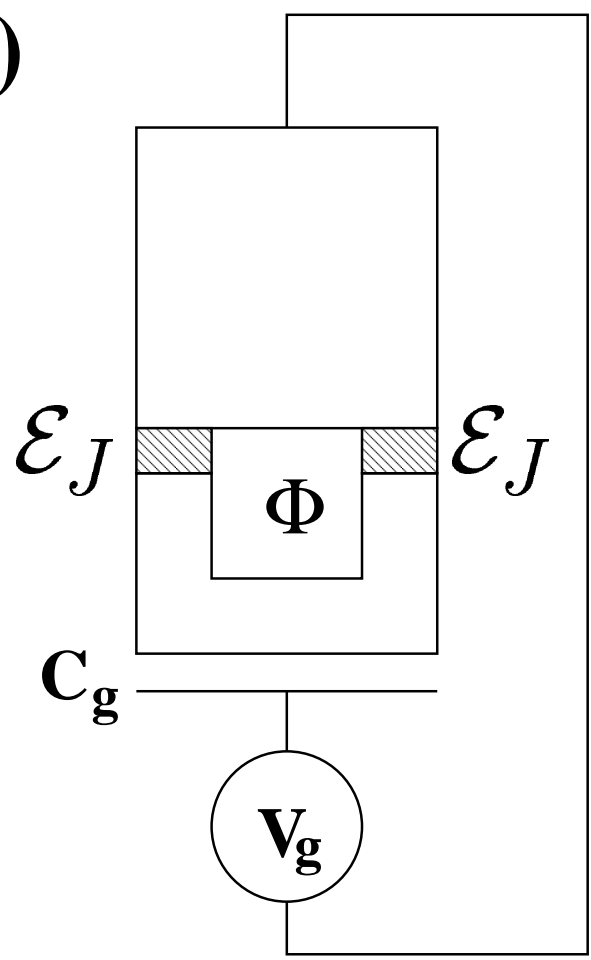
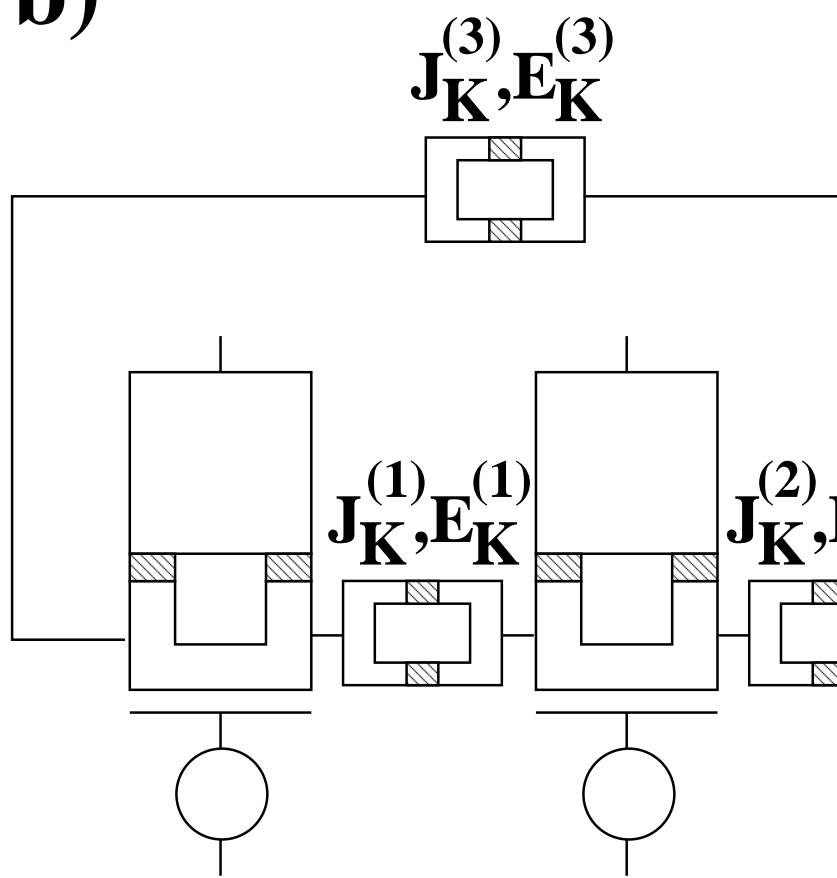
\end{document}